\documentclass[sigplan]{acmart}

\usepackage{multirow}
\usepackage{graphicx}
\usepackage{caption}
\usepackage{subcaption}
\usepackage{balance} 
\usepackage{listings}

\newcommand{\reffig}[1]{Fig.~\ref{#1}}
\newcommand{\reftab}[1]{Table~\ref{#1}}
\newcommand{\refsec}[1]{Section~\ref{#1}}

\usepackage[defaultlines=3,all]{nowidow}
\usepackage{enumitem}
\setlist{nolistsep}
\setlist{nosep}

\begin{document}

\title{Towards Diversity-tolerant RDF-stores}


\author{Masoud Salehpour}
\affiliation{\institution{University of Sydney}}

\author{Joseph G. Davis}
\affiliation{\institution{University of Sydney}}

\renewcommand{\shortauthors}{Masoud Salehpour and Joseph G. Davis}

\begin{abstract}
We propose a novel approach to designing RDF-stores with the goal of improving the consistency and predictability of query performance. When designing these systems, three properties are commonly desired: support for the full range of SPARQL query features (Q), support for widely varying RDF datasets in terms of structuredness and size (S), and high performance (P). We develop the empirical SPQ conjecture which states that it may be impossible to achieve all the three desiderata simultaneously. We present a strong case for its plausibility based on our experimental results. The tradeoffs among the three and design guidelines based on the SPQ conjecture are also discussed. \end{abstract}


\maketitle

\section{Introduction}

\begin{figure}[t]
\centering
\resizebox{\linewidth}{!}{%
\includegraphics{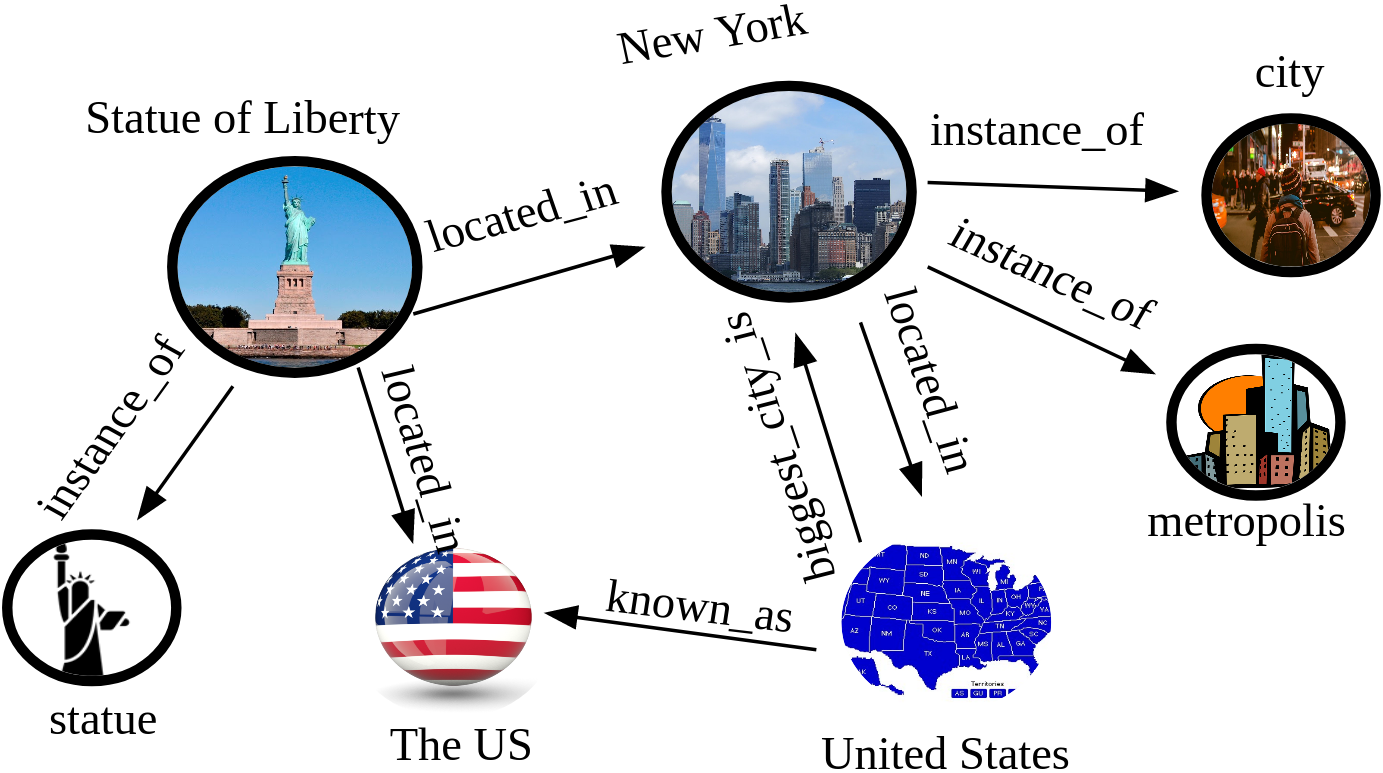}
}
\caption{A simple Knowledge Graph describing \mbox{Statue of Liberty}.}
\label{fig::kg}
\end{figure}

Knowledge Graphs (KGs) are large collections of information about real-world entities (such as people, places, organizations, movies, books, music albums, proteins, genes, drugs, etc.) and their interconnections~\cite{weikum-fnts,hogan2021knowledge-survey}. Many KGs have been deployed both by private enterprises and in the public domain over the last ten years such as \textit{Google Knowledge Graph}~\cite{google-kg} as a well-known example of a private KG alongside \textit{BabelNet}\footnote{Available Online: \url{http://babelnet.org}} as a salient example of a KG with publicly accessible contents. KGs are typically represented using the RDF (Resource Description Framework) data model which is a directed, labeled graph-like structure for representing the content of a KG using a set of triples of the form <subject predicate object>. \reffig{fig::kg} depicts an example of a KG to describe the fact that the \mbox{``Statue of Liberty''} is a heritage site located in \mbox{``New York''}. In this figure, ``Statue of Liberty'' is a subject, ``located\_in'' is a predicate, and ``New York'' is an object. The content of this KG can be represented by the following triples\footnote{We use human-readable names in this paper. However, Internationalized Resource Identifiers (IRIs) are generally used to represent triples.}:

\begin{figure}[h]
    \centering
    \begin{lstlisting}[basicstyle=\scriptsize\ttfamily\color{black},tabsize=2,showstringspaces=false,numberstyle=\small\tt\color{gray}]
    StatueOfLiberty  located_in        NewYork
    StatueOfLiberty  located_in        The US
    StatueOfLiberty  instance_of       statue
    NewYork          instance_of       city
    NewYork          located_in        UnitedStates
    NewYork          instance_of       metropolis    
    UnitedStates     known_as          The US
    UnitedStates     biggest_city_is   NewYork
 \end{lstlisting}
    \label{fig:triples-KG-ny}
\end{figure}


The widespread adoption of KGs for the representation and exchange of semantic data has highlighted the need for employing efficient \textit{RDF-stores}, i.e., customized data management systems for KGs represented using the RDF data model. A number of RDF-stores have been developed over the past few years. Early approaches such as~\cite{Dan2007,DanSW} proposed different ways of adapting relational databases for RDF data management such as grouping triples with the same predicate name into property tables, mapping these onto a column store, and creating materialized views for frequent joins. The high number of resulting joins imposed restrictions on the performance of these systems~\cite{AntiDan}.

Over time, a range of new architectures for storing, indexing, and querying have also been developed such as gStore~\cite{gStore} using a disk-based adjacency list tables to store KGs and executes queries by mapping them to a subgraph matching task and~\cite{Turbo} which treated RDF graphs as labeled graphs and applied subgraph homomorphism methods for query processing.~\cite{RDF3x} proposed ``RISC-style'' architecture to leverage multiple query processing algorithms and optimization.~\cite{tamer2019} developed a workload-adaptive and self-tuning RDF-store using physical clustering of the underlying data. Also, a variety of popular, transactional RDF-stores have been developed such as OpenLink Virtuoso\footnote{\url{https://virtuoso.openlinksw.com/}}, Blazegraph\footnote{\url{https://blazegraph.com/}}, Jena-TDB2\footnote{\url{https://jena.apache.org/}}, among others. In these and similar systems, the designers have exploited a large number of technical choices and architecture including several exhaustive indexing strategies~\cite{Thomasindex,RDF3x}, complicated cardinality estimation, compression techniques, and dictionary encoding (to keep space requirements reasonable for excessive indexing) as well as vectored execution, compiler-based early evaluation of query expressions (i.e., data-independent sub-queries), and run-time join type selection (e.g., merge or hash joins), to name a few. Details of these systems are discussed in comprehensive surveys such as~\cite{Kaoudi,Ozsu,Survey2018,SurveySPARQL,saleem}. Despite these efforts, the RDF-store physical design and query optimization continue to be an open problem. There have been extensive reports of a good portion of queries being timed out or having to be reformulated~\cite{watdiv}.

It appears that the inconsistency in query performance is at least in part attributable to the characteristics of the RDF datasets and the diversity of SPARQL query features. More precisely, managing large-scale RDF data poses the efficiency challenges for the storage layout and query processing: (i) RDF is designed as a schema-relaxable (sometimes referred to as schema-free or schema-last) representation format in which a global schema is absent along with the disparity of predicate names across instances (of the same type), (ii) non-locality of RDF data makes query optimization efforts expensive (non-locality comes with the nature of RDF data since it follows a fine-grained model in which ``triple'' is the only unit of information as compared to ``record'' in RDBMS in which both an entity's attributes and relationships to other entities are available), and (iii) no \mbox{a-priori} assumption about the structure of queries can be made because of the widely varying (SPARQL) query features~\cite{bonifatibook}, including required and optional graph patterns, to name a few. The evolving structure of the data and the widely varying (SPARQL) query features can easily render the RDF data management problem into a Sisyphean task~\cite{RDF3x}.

In this paper, we propose the \textit{SPQ conjecture} which states that we can simultaneously achieve only any two of: (i) \textit{support for the full range of SPARQL query features} (\textbf{Q}), (ii) \textit{support for widely varying RDF datasets (in terms of structuredness and size)} (\textbf{S}), and (iii) \textit{high performance} (\textbf{P}). An earlier version of this appeared as part of a credible benchmarking effort~\cite{watdiv} which highlighted the difficulties in achieving consistent performance across varying query features. In this conjecture: \textbf{Q} includes required and optional graph patterns, aggregation, subqueries, negation, along with their conjunctions and disjunctions as well as creating values by expressions and extensible filtering. \textbf{S} refers to the structuredness and size of RDF data which is introduced and defined in~\cite{IBMapple} as follows: Each dataset comes with a number of types and corresponding entities or instances of each type.\footnote{For instance, an academic dataset may contain different types such as ``Professors'', ``Students'', and ``Courses''. It then may have ``Professor0'', ``Student101'', and ``COMP5050'' as instances of these types.} The structuredness is a composite metric that characterizes each dataset using the number of types, properties, properties per type, instance entities per type, and instance properties for these entities. Intuitively, the structuredness of a dataset D with respect to a type T is determined by how well the instance data in D conform to type T. If each entity in D sets values for most (if not all) of the properties of T, then all the entities in D have a similar structure that conforms to T. In this case, we say that D has high structuredness with respect to T. The structuredness of T is affected by the sparsity or absence of its properties across instances. Any RDF dataset's level of structuredness can be quantified by a normalized value in the [0, 1] interval, with values close to 0 corresponding to low structuredness, and 1 corresponding to perfect structuredness. In a KG dataset with high structuredness, we expect that for any two instances of the same type, the instances have exactly the same properties. \textbf{P} is measured by the time it takes for the correct result to be returned. It is difficult to prove the SPQ conjecture formally at this stage. We, however, present empirical evidence through a set of experiments, the results of which are included in this paper. It is hoped that this conjecture can contribute to the ongoing debate on the tradeoffs relating to efficient support for the full spectrum of SPARQL queries against large-scale KG datasets with varying levels of structuredness. This research is inspired by Brewer's CAP Theorem~\cite{cap1,cap2,harvest}, \mbox{Stonebraker et. al.}~\cite{oneSize,onesize2}, and HAT Systems~\cite{hat} which can be paraphrased in this context to relate KG structuredness (\textbf{S}) and high performance (\textbf{P}) to query features (\textbf{Q}).

Our experiments focus on query performance using major open-source and publicly available systems, namely, Virtuoso, Blazegraph, and Apache Jena-TDB2 delivering the following services (i) disk-oriented persistent model with disk-resident indexes, (ii) buffer-pooling (and multi-threading support), (iii) full support for read-only queries (although we explicitly focus on read-only queries, most of the findings are also applicable to update queries since in most RDF-stores, update queries usually include a search predicate to determine the triples are supposed to be modified where typical optimization and execution techniques of read-only queries also apply to the search phase of update queries; the update procedure itself can be either deferred until a second phase or merged into the search phase if it is not \textit{dangerous}), and (iv) single-node and non-failing databases. 

Our contributions in this paper include:

\begin{itemize}

 \item We develop \textit{SPQ conjecture} to state that it is not plausible to design high-performance RDF-stores supporting a variety of query features against diverse RDF datasets. However, by explicitly handling these properties, designers can shift their attention to balancing the tradeoffs.
 
 \item We present some evidence to support our conjecture by performing experiments using well-known benchmark datasets (with varying levels of structuredness and size) and analyze representative queries employing major RDF-stores. 
 
 \item We classify the SPARQL query spectrum to High Performance Plausible (HP-Plausible) and High Performance Dubious (HP-Dubious) where RDF-stores guarantee efficient execution of HP-Plausible queries.
 
\end{itemize}

The remainder of this paper is organized as follows. In~\refsec{sec:spq}, we discuss the impacts of S and Q on P. This includes analyses of the tradeoffs among these three.~\refsec{sec:experiments} presents our experimental setup including the KG benchmark characteristics, computational environment, and system settings. The summary of our experiments as the evidence for the SPQ conjecture is also included as well as a bifurcation of the spectrum of SPARQL queries, namely, HP-Plausible and HP-Dubious, and their performance characteristics. We present our conclusions and future work in~\refsec{sec:conclusion}.


\section{SPQ Conjecture}
\label{sec:spq}
In this section, we provide brief background information on KG query processing. We also explain how S and Q can influence performance (P) by negatively affecting the accuracy of cardinality estimation, the effectiveness of indexes, and plan enumeration. Tradeoffs among the three are also analyzed.

\subsection{KG Query Processing}

In general, KG query processing proceeds as follows. The KGs are first loaded into an RDF-store, triggering the creation of related indexes which are crucial to performance. Major RDF-stores typically maintain a set of six indexes, known as exhaustive indexing, which cover all possible triple orders~\cite{survey-native-non-native}. After the KG and all indexes are loaded, the users are able to submit their queries. The RDF-store parses a given query according to the query language grammar (i.e., SPARQL) and generates the corresponding syntax tree for the query. During this phase, the RDF-store also validates that the user has appropriate permissions to run the query. The generated syntax tree is then transformed into a logical operator graph as the next phase. This is followed by logical operator graph optimization. For example, the execution order of different parts of a query such as filtering or sorting should be planned. The next phase is to choose the best implementations for each logical operator referred to as physical optimization. This is a complex process whose outcome is to generate a low-cost \textit{physical operator graph} to perform the query. Finally, the execution phase is performed, wherein the desired data items are retrieved and the answer to the query will be returned~\cite{svenbook}.

SPARQL enables users to formulate their queries by specifying \textit{what} is desired as an answer without specifying (in any way) \textit{how} the answer is to be retrieved. This is often referred to as the \textit{non-procedurality} feature of SPARQL. On this basis, the user is not required to be involved in specifying the step-by-step access plan to execute the query since it is generally feasible to leave this to the employed RDF-store where the RDF-store's query optimizer is expected to build plans using their indexing schemes to execute queries as efficiently as possible. A seemingly natural way of executing queries faster is to reduce the extent to which the underlying data is touched by selecting suitable indexes and establishing the ordering by which intermediary results can be joined with the least overall cost. To this end, query optimizers are usually designed to pick joins in ascending order of selectivity (otherwise, too many triples are needed to be read). Theoretically, as long as the cardinality estimations (and the cost model) are accurate, RDF-stores can construct the optimal query plan and subsequently execute queries efficiently. However, S and Q can have major impacts on P as explained below.

\subsection{Impacts of S and Q on P}
 
\noindent \textbf{Impacts of Structuredness (S) on Performance (P)}. The performance of RDF-stores relies on finding the least-cost execution plan where estimated cardinalities have a significant impact on the optimal plan generation. Computing cardinalities for datasets with lower structuredness is challenging since it implies that triples are more heterogeneous (i.e., datasets with a larger diversity of predicate names) and it is unlikely that existing data structures are compact enough to group all correlated triples given the heterogeneity of such a dataset. Lower structuredness fundamentally complicates the query optimizers’ efforts for deriving reliable estimates.

While lower values for structuredness typically affect the accuracy of cardinality estimations, higher values can decrease the effectiveness of the indexes. More specifically, higher values of structuredness typically indicate the presence of identical properties across triples which impact the density of indexes negatively since density usually shows the ratio of unique values. Higher the density lower the selectivity and subsequently lower P for highly structured datasets.

\noindent \textbf{Impacts of Query features (Q) of Performance (P)}. Queries are sets of triple patterns requiring pattern matching. They include variables that imply joins. Given a query with a number of triple patterns, a join of two triple patterns in an RDF-store can be performed as follows. It begins with an index lookup to retrieve all triples that match with the first triple pattern of the query. During this step, matched triples are usually read from the disk and cached in memory; often referred to as the \textit{left} side, or the outer side, of the join. A similar lookup is to be performed to retrieve triples of the \textit{right}, or the inner, side of the join.
When both sides (i.e., left and right) are retrieved, the RDF-store uses a join algorithm to combine them to answer the query. The choice of join algorithm can significantly affect the efficiency of the join processing. The number of algorithms that can be used by an RDF-store for joining triple patterns is an important factor. Typical algorithms are \textit{nested loop join}, \textit{hash join}, and \textit{merge join}. Each of these algorithms shows performance advantages in situations that can arise in performing a join between two triple patterns. For instance, the nested loop join is typically efficient when only a ``\textit{small}'' number of triples qualify from the outer side of the join, or when the inner side of the join is ``\textit{small enough}'' that all index and data become resident in memory during the join processing. The merge join (sometimes referred to as \textit{merge scan join} or \textit{sort merge}) is more likely to be efficient when both sides are already sorted~\cite{onielBook}. Although the importance of using an efficient join algorithm is recognized, choosing the most optimal join algorithm to process a given query is still a major challenge.

Joins of three or even more triple patterns are performed in a binary way, where two triple patterns are joined at a time. The outcome of the first two joins is typically stored in memory as an intermediate result, which is then used to join with the third, and is repeated for any further triples. Due to the \textit{non-procedurality} feature of KG queries, the order of joins is not determined by users and it is left entirely up to the RDF-store. Typically, the RDF-store begins by enumerating all possible orderings and selecting the most efficient one. However, the major challenge is that this requires accurate statistics and a great deal of computational effort for queries containing multiple joins.

Generating optimal execution plans can lead to faster query execution times. However, it is obvious that the higher the number of joins, the higher the number of possible orderings which can make the index choice and plan enumerations difficult. The number and complexity of query features would make the search space larger in order to find an efficient plan. The search space can consist of $n!k^n$ plans where $n$ denotes the number of required joins and $k$ the number of algorithms for the join processing that are implemented by the RDF-store (like merge join and hash join)~\cite{rox-phd-thesis}. For instance, one of the difficulties in executing queries with conjunctions or disjunctions such as union and optional patterns is that RDF-stores need to generate plans with all possible combinations~\cite{linkedBook-harth2014,c5-linkedBook-harth2014}. This is obviously expensive since cardinality estimations are less likely to be helpful in plan generation. \footnote{In theory, sampling-based query execution may help in these cases, but it is not very practical to follow a sampling-based plan enumeration (or a randomized selection) to find the optimal plan since for many queries, the sample would have to be large to be useful which is obviously very expensive and itself affect the performance negatively.} As a result, RDF-stores are unlikely to generate optimal plans for queries with multiple joins as well as optional patterns and unions.


\subsection{SPQ Tradeoffs}
SPQ conjecture can make explicit the tradeoffs in designing RDF-stores. For every pair QP, QS, PS, it can elucidate the potential effects on the third property. We present some examples of each pairing in this section. These examples suggest some range of systems and tradeoffs in which it is clearly impossible to satisfy both Q and S simultaneously, one or both must be relaxed to some extent depending on the application requirements. P is highly desirable in many applications. Therefore, we briefly discuss trading either Q or S off for P.

\noindent\textbf{Trading Q off for SP.} Relaxing Q is typically straightforward in which some certain query features are sacrificed. RDF-stores, therefore, guarantee little or no latency to execute remaining query features, as a side effect, the database user's wait for the result (often the last result item) is negligible. We refer to these queries as HP-Plausible (see \refsec{sec::result} for details). As trading Q off for SP, we suggest the support for queries that only contain a set of triple patterns called a basic graph pattern which are bounded (no language tags are included as well).\footnote{Basic graph patterns match a subgraph of the KG data. The operator ``and'' is usually used along with the set of basic triple patterns.} This tradeoff seems suitable for many web-based applications.

\noindent\textbf{Trading S off for QP.} We present two examples for this tradeoff. First, RDF-stores can guarantee P just for certain levels of structuredness (S). It is possible by designing fixed, locality-aware physical layouts (e.g., clustering triples into records) with regard to the levels of structuredness that are intended to be supported. This tradeoff enables easy customization of the physical data structures and indexes in the database based on the structuredness, resulting in more efficient I/O and cache utilization (through minimizing dirty cache pages, virtual memory swapping warning, disk random read), better indexing and data localization, and fewer intermediate tuples during query evaluation. Second, RDF-stores can maintain QP, but reduce the fraction of data reflected in the response, i.e. the completeness of the answer to the query. Some applications obviously do not tolerate this tradeoff because any deviation from the completeness renders the result useless, but in some applications, RDF-stores can return incomplete results for avoiding long query optimization and execution that looks unlikely to be performed efficiently given the size and structuredness of underlying dataset.

\section{Experimental Setup}
\label{sec:experiments}
In this section, we report our experimental context and details of the KG benchmark datasets used for the experiments. This also includes details of our computational environment and RDF-stores. The summary of our experiments as the evidence for the SPQ conjecture is also presented.

\begin{table*}[t]
\centering
\begin{tabular}{clrcrrc}
\toprule
 Benchmark & Scale (nominal) & \#Subjects & \#Predicates & \#Objects & \#Triples & \#Structuredness \\
 \toprule\toprule
\multicolumn{1}{ c }{\multirow{3}{*}{WatDiv} } &
\multicolumn{1}{ c }{100K} & 5,597 & 86 & 13,337 & 10,6250 & 0.53 \\
\multicolumn{1}{ c }{} &
\multicolumn{1}{ c }{10M} & 5,212,385 & 86 & 9,753,266 & 108,997,714 & 0.46\\ 
\multicolumn{1}{ c }{} &
\multicolumn{1}{ c }{1000M} & 52,120,385 & 86 & 92,220,397 & 1,092,155,948 & 0.46\\ \cline{1-7}
\hline
\hline
\multicolumn{1}{ c }{\multirow{2}{*}{SP2Bench} } &
\multicolumn{1}{ c }{100K} & 19,368 & 58 & 50,265 & 100,071 & 0.73\\ 
\multicolumn{1}{ c }{} &
\multicolumn{1}{ c }{100M} & 17,823,522 & 78 & 23,516,800 & 100,000,374 & 0.76\\ 

\hline
\hline
\multicolumn{1}{ c }{\multirow{3}{*}{LUBM} } &
\multicolumn{1}{ c }{100K} & 17,174 & 17 & 13,946 & 100,543 & 0.96\\
\multicolumn{1}{ c }{} &
\multicolumn{1}{ c }{100M} & 21,735,127 & 17 & 16,156,825 & 138,318,414 & 0.95\\ 
\multicolumn{1}{ c }{} &
\multicolumn{1}{ c }{1000M} & 217,206,844 & 17 & 161,413,041 & 1,335,081,176 & 0.95\\ \cline{1-7}

\end{tabular}
\caption{Statistics of the benchmark datasets. Structuredness values are normalized in the [0, 1] interval with values close to 0 corresponding to low structuredness, and 1 corresponding to high structuredness.}
\label{table::kgs}
\end{table*}
\subsection{Datasets}
We used three well-known benchmarks in this research. These are publicly available datasets along with a collection of queries (over 50 queries). These benchmarks are: Waterloo SPARQL Diversity Test Suite (WatDiv)~\cite{watdiv}, SPARQL Performance Benchmark (SP2Bench)~\cite{sp2b}, and Lehigh University Benchmark (LUBM)~\cite{lubm}. These benchmarks follow specific rules that allow us to scale the datasets to arbitrary sizes using their scale factors. \reftab{table::kgs} shows the statistical information related to the benchmarks. The RDF representations of these benchmarks are available in different formats such as Turtle and XML. We used the RDF/N-Triples format. This is a well-known and popular subset of turtle in which each individual line of the file should contain all necessary information to parse the triple on that line independent of the rest of the document. This is perhaps the reason behind its popularity.~\cite{linkedBook-harth2014,c1-linkedBook-harth2014}.

\subsection{Computing Structuredness}
We followed the instructions of~\cite{IBMapple} to compute the structuredness which the first step is clearly the determination of the type system of a dataset. Conceivably, one might infer the type system T of a dataset D intensionally, through an RDFS or OWL specification associated with D. In practice, however, many datasets do not come with such specifications. To address the fact that often there is no schema available in datasets to apply an intensional approach, we determine the type system T of D extensionally through the dataset itself, and thus do not need any schema-level information. Specifically, we scan D looking for triples whose property is <http://www.w3.org/1999/02/22-rdf-syntax-ns\#type>, and for these triples, we extract the type T that appears as the object of the triple. We then determine the properties of a type T through the union of all the properties that the instances of type T have. We then count the number of entities for which property p has its value set in the instances (we are only interested in whether a property is present or not for a particular entity rather than how many times the property is set for this entity). The structuredness can characterize each dataset using the number of types, properties, properties per type, instance entities per type, and instance properties for these entities.

In our experiments, we need datasets with the same size but varying structuredness and a set of queries that return the same results for each dataset. To this end, we developed a generator that accepts as input any dataset (like a dataset generated from any of the existing benchmarks) along with the desired level of structuredness and size, and uses the input dataset as a seed to produce a dataset with the desired size and structuredness. Our data generator offers complete control over both the structuredness and the size of the generated data. There is an interaction between data size and structuredness such that altering the size of a dataset can affect its structuredness, and correspondingly altering the structuredness of a dataset can affect its size. We therefore could not just randomly add/remove triples in the input dataset until we reach the desired output size. As a solution, our generator comes in the form of two objective functions, one for structuredness and one for size, we then formulate and solve these objective functions using existing integer programming solvers. We borrowed this idea of using integer programming from~\cite{IBMapple} in which interested readers may find more details about the objective functions and the corresponding constraints.

\begin{figure*}[htp]
\centering
\includegraphics[width=.33\textwidth]{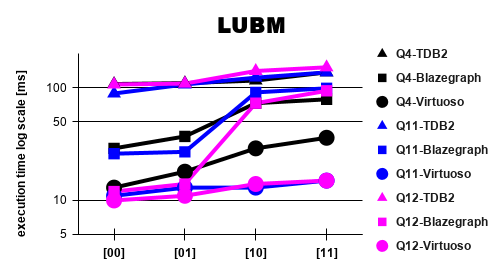}\hfill
\includegraphics[width=.33\textwidth]{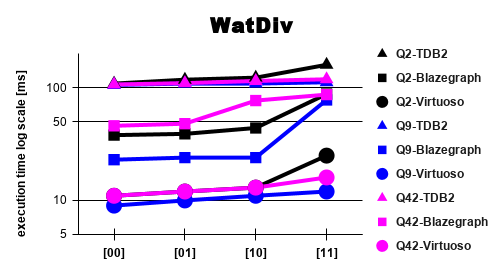}\hfill
\includegraphics[width=.33\textwidth]{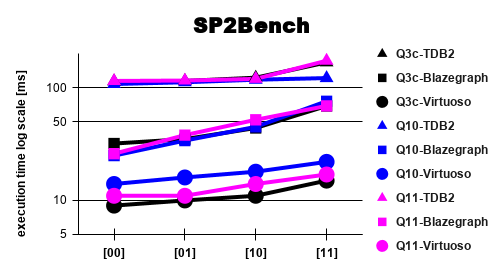}
\caption{Queries that only contain a set of bounded triple patterns (no language tags are included as well) are executed across datasets with different sizes and levels of structuredness employing different RDF-stores. $X$ axis represents different combinations of size and structuredness values where \textit{``[00]''} means low size and low structuredness, \textit{``[01]''} represents low size and high structuredness, \textit{``[10]''} is for high size with low structuredness, and finally \textit{``[11]''} means high size and high structuredness. In this figure, the low size means datasets with around 100K triples while the high size refers to datasets with over 1000M triples combined with low structuredness which means 10\% of original datasets' structuredness and high size means 90\% of each dataset's original structuredness.}
\label{fig::sp}
\end{figure*}

We implemented our algorithm in C and performed experiments on the datasets.\footnote{The source code and data are available through: \url{https://github.com/m-salehpour/SPQ}}. We used \textit{lpsolve 5.5}\footnote{\url{http://lpsolve.sourceforge.net/5.5/}} as our integer programming solver with a timeout of 120 seconds per problem. We generated 9 datasets of different structuredness (ranging from 0.1 to 0.9 percentage of the original structuredness) and the same size (the final size of each generated dataset is around 70\% of the original dataset size). For instance, we generated 9 datasets of different structuredness and the same size using WatDiv100K. Similarly, 9 datasets are generated using WatDiv10M or 9 datasets using LUBM100K, etc. To prepare the datasets, we had to process over 60 billion triples from our input datasets, and we generated over 6TB of intermediate files to prepare the final output files. Notice that in some cases the integer programming solver could not find an optimal solution or in some other cases the solver finds a solution that has the right structuredness but misses the size. As suggested in~\cite{IBMapple}, we performed post-processing to adjust the size as close as possible, e.g., by attempting to remove from triples with the same subject and predicate.\footnote{All needed post-processing attempts are explained in detail in~\cite{IBMapple}.}

\subsection{System Settings}
\noindent \textbf{Computational Environment.} To speed up the data preparation process, we used four Virtual Machines (VMs) with identical configurations in parallel: 2.3GHz AMD Processor, running Ubuntu Linux (kernel version: 5.4.0-40-generic), with 64GB of main memory, and 16 vcores, but we selected just one of them as our benchmark system for actual query processing with 512K L2 cache and 8TB instance storage capacity. The VM cache read is roughly 2639.65MB/sec and the buffer read is roughly 32.97MB/sec (i.e., the output of the ``hdparm -Tt'' Linux command). The operating system is set with almost no ``soft/hard'' limit on the file size, CPU time, virtual memory, locked-in-memory size, open files, processes/threads, and memory size using Linux ``ulimit'' settings.

\noindent \textbf{RDF-stores.} We chose three different RDF-stores: (1) column-store Virtuoso (Open Source Edition, Version 07.20.3230--commit 4a668a5), (2) Blazegraph\footnote{Previously known as Bigdata DB.} (Open Source Edition, version 2.1.6--commit 6b0c93), and (3) Apache Jena-TDB2 (version 3.16.0). We selected these RDF-stores since these are open-source, publicly available RDF-stores among the Top 10 ranking list of DB-Engines website.\footnote{\url{https://db-engines.com/en/ranking/rdf+store}} We tuned them based on their official documentations.

\begin{figure*}[htp]
\centering
\includegraphics[width=.33\textwidth]{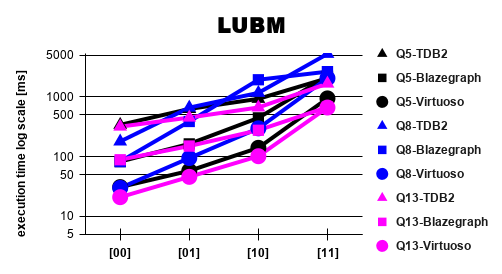}\hfill
\includegraphics[width=.33\textwidth]{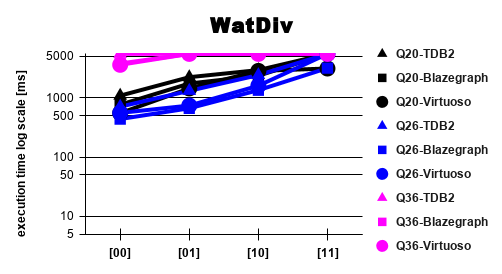}\hfill
\includegraphics[width=.33\textwidth]{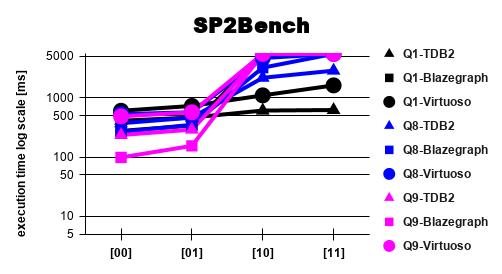}
\caption{The trends show that achieving high performance is mutually conflicting with higher values of size and structuredness when Q in SPQ conjecture is picked to the extent that a high proportion of the queries time out (i.e., exceeding 5000ms).}
\label{fig::sORp}
\end{figure*}

\subsection{Results}
\label{sec::result}
As part of the experimental evaluation, we measured the query execution time. This is an end-to-end time computed from the time of query submission to the time when the result is outputted. After the execution of each query, we carefully checked to ensure that the output results are correct and exactly the same across different structuredness. For fairness, the times reported for each RDF-store are averaged\footnote{Geometric mean is used.} over 5 successive runs (with almost no delay in between). Due to space constraints, we present the summary of our experiments in the following as evidence for our conjecture, but the full datasets, algorithms, and results are available online.\footnote{The source code and data are available through: \url{https://github.com/m-salehpour/SPQ}}

\textbf{HP-Plausible.} We identify HP-Plausible queries in our experiments to support the conjecture that scarifying some query features will increase the likelihood of delivering high performance. In the context of SPQ conjecture, it means picking SP and relaxing Q. There is a supporting sign which is shown in \reffig{fig::sp} where SPARQL queries that only contain a set of bounded triple patterns (no language tags are included as well) are executed with high performance across different sizes and spectrum of structuredness employing different RDF-stores. In the figure, $X$ axis represents different levels of size and structuredness values where ``[00]'' means the low size and low structuredness, ``[01]'' represents the low size and high structuredness, ``[10]'' is for the high size with low structuredness, and finally ``[11]'' means the high size and high structuredness. In our experiments, the low size means datasets with around 100K triples while the high size refers to datasets with over 1000M triples combined with low structuredness which means 10\% of original datasets' structuredness and high size means 90\% of each dataset's original structuredness. The $Y$ axis shows the execution times of queries. In \reffig{fig::sp}, titles show the name of the corresponding dataset along with colors that are used to distinguish different queries while employed RDF-stores are shown by different point shapes (i.e., circle, square, and triangle to refer to Virtuoso, Blazegraph, and Jena-TDB2, respectively). For instance, the leftmost sub-figure of \reffig{fig::sp} shows LUBM-Q12 are milliseconds against [00] combination (i.e., LUBM-100K with 10\% of the original structuredness) using Virtuoso. The trend in~\reffig{fig::sp}, offers a pragmatic takeaway to support our SPQ conjecture that by picking SP and relaxing Q (i.e., scarifying some query features like sequential modifiers and conjunctive optional patterns), the likelihood of achieving high performance across datasets with different structuredness and size values increases considerably. Notice that our aim is not to compare RDF-stores, but we are interested to highlight that each HP-Plausible individual query is often executed in the same order of milliseconds even though in the presence of a varying spectrum of structuredness and scale values.

\textbf{HP-Dubious.} SPQ conjecture suggests that S and P are mutually conflicting when Q is picked (SPARQL query features). Intuitively, it means that high performance is not guaranteed in the event of a dataset with a large size and high structuredness in the presence of many SPARQL queries (e.g., optional matches including property paths and sequential modifiers as well as unions of conjunctive queries). \reffig{fig::sORp} illustrates the conflict with S and P for many query features. In this figure, the $X$ and $Y$ again represent different combinations of size and structuredness values along with the execution times of queries distinguished by colors using RDF-stores shown by different point shapes. For instance, the sub-figure which is shown in the middle of \reffig{fig::sORp} shows WatDiv-Q26 is executed in around 1 second against lower size and structuredness values while its execution time exceeds 5 seconds as values of size and structuredness are increased (please note that 5000 seconds here represents the concept of time-out). One credible upshot from the trends in~\reffig{fig::sORp} is that achieving high performance is mutually conflicting with higher values of size and structuredness when Q in SPQ conjecture is picked to the extent that in our brief experiments a high proportion of the queries time out.

Based on our experiments, we extend the notion of performance for querying RDF datasets into two performance models, namely, HP-Plausible and HP-Plausible. This helps us to classify which among the wide array of structuredness and queries are achievable with high performance. Our results suggested that RDF-stores guarantee little or no latency to execute HP-Plausible, but a broad range of queries are not included in these groups such as queries consisting of a set of at least two node-disjoint paths. In addition, the performance can be degraded considerably when it is inevitable to join intermediary results of a query that do not fit the small-but-fast memory (i.e., the CPU cache) or even worse when the size of intermediary results exceeds the RAM buffer size and pipelining is not possible. RDF-stores then have to switch to scalable I/O-based merge solutions (e.g., sort-merge-join or hybrid hash-join) to return the result.

As we summarized in~\reftab{table::summary}, a range of queries are HP-Plausible. HP-Dubious contains a wider range of queries that can be high performance when the datasets have lower levels of structuredness and sizes. However, achieving high performance for HP-Dubious cannot be guaranteed against highly structured datasets at scale.
 
\begin{table}[htp]
\centering
\begin{tabular}{ ll } 
\hline
HP-Plausible & Single Match (SM) \\
& Bounded Multiple Match (BMM)\\ 
& Property Path of length 1 (P1) \\ 
\hline
\hline
HP-Dubious & Graph Patterns (GP)\\
& Aggregations (AG) \\
& OPTional patterns (OPT)\\ 
& Union (U) \\ 
& Arbitrary Property Path (APP) \\
& Sequential Modifiers (MO) \\
\hline
\end{tabular}
\caption{Summary of query performance models considered in this paper.}
\label{table::summary}
\end{table}

\section{Conclusion}
\label{sec:conclusion}
We have introduced the SPQ conjecture and explained some of the tradeoffs implied by this. We have presented experimental evidence in support of the conjecture. We also developed a bifurcation of the spectrum of SPARQL queries, namely, HP-Plausible and HP-Dubious, and their performance characteristics. As part of future work, we propose to design the next generation of the RDF-stores which explicitly incorporate these tradeoffs derived from the SPQ conjecture with the goal of achieving higher consistency and predictability of RDF query performance.

\balance

\bibliographystyle{ACM-Reference-Format}
\bibliography{main}


\end{document}